\documentclass[twocolumn]{aastex631}

\usepackage{amsmath}

\shorttitle{gravity eigen-modes}
\shortauthors{M. Farhang and N. Khosravi}
\begin{document}
\title{ Reconstruction of A Scale--Dependent Gravitational Phase Transition}

\author{Marzieh Farhang}
\email{m\_farhang@sbu.ac.ir}
\affiliation{\scriptsize {Department of Physics, Shahid Beheshti University, 1983969411, Tehran, Iran}\\}

\author{Nima Khosravi}
\affiliation{\scriptsize {Department of Physics, Shahid Beheshti University, 1983969411, Tehran, Iran}\\}
\email{n-khosravi@sbu.ac.ir}

\begin{abstract}
In this work we extend our earlier phenomenological model for a gravitational phase transition (GPT) \citep{Farhang:2020sij}
and its generalization to early times \citep{Khosravi:2021csn}
by letting the modifications in the linearly--perturbed Einstein equations be scale-dependent. 
These modifications are characterized as deviations of the parameters $\mu(z,k)$ and $\gamma(z,k)$ from their values in general relativity (GR). 
The scale-dependent amplitudes of modified $\mu(z,k)$ and $\gamma(z,k)$ and the parameters defining the phase transition, along with  the standard cosmological parameters, are measured by various data combinations. 
Out of the perturbation parameters, we construct gravity eigenmodes which represent patterns of perturbations best detectable by data. 
We detect no significant deviation from GR in these parameters. However, the larger parameter space produced due to the new degrees of freedom allows for 
the reconciliation of  various datasets which are in tension in $\Lambda$CDM.
In particular, we find $H_0=71.9\pm 9.2$ from anisotropies  of the Cosmic Microwave Background as measured by {\it Planck}  \citep{pl18} 
and various measurements of the Baryonic Acoustic Oscillations, in agreement with  local Hubble measurements \citep{Riess:2021jrx}. We also find that the $\sigma_8$ tension between the measurements of Dark Energy Survey \citep{Kazin:2014qga} and  {\it Planck} is reduced to less than  $1\sigma$. 

\end{abstract}

\section{Introduction} \label{sec:intro}
 The standard model of cosmology, aka the $\Lambda$CDM scenario, in spite of its incredible achievements in explaining the observed sky,
suffers from serious limitations. Certain fundamental questions are left unanswered in this scenario such as the nature of dark matter and  dark energy and the cosmological constant problem. 
Moreover, the model has been challenged by observations in recent years.
 Among the most severe is the inconsistency of the Hubble constant as directly measured by local data \citep{Riess:2021jrx,Murakami:2023xuy} and $\Lambda$CDM-based inferences from the Cosmic Microwave Background (CMB) anisotropies \citep{pl18}.
 The Hubble tension is accompanied by milder ones including the $\sigma_8$ tension \citep{Abbott:2017wau,hey20}, the deviation of the measured CMB lensing amplitude from unity, the parameter inconsistencies  in the  CMB low and high $\ell$ multipoles, and the CMB spatial anomalies \citep{pl18}. 
 All these anomalies can be due to  systematics in  observational and analysis pipeline.
 In this paper, however, 
 we follow the viewpoint that the tensions might be due to theoretical insufficiencies. We mainly focus on $H_0$ and $\sigma_8$ tensions here.
 
In the attempts to address the $H_0$ tension through modifications to the $\Lambda$CDM model,  a very natural place to look for is  dark energy. See, e.g., \cite{DiValentino:2017iww,Zhao:2017cud,Khosravi:2017hfi,yan18,Yang:2018uae,Banihashemi:2018has,Li:2019yem,kee19,Raveri:2019mxg,Yan:2019gbw,dival20,Braglia:2020iik,Gomez-Valent:2020mqn,luc20,DiValentino:2019ffd,DiValentino:2019jae,Banihashemi:2018oxo,Banihashemi:2020wtb} 
and for  a comprehensive review see \cite{DiValentino:2021izs}. 
Modifications in the early universe are also able to lessen the $H_0$ tension by reducing the sound horizon $r_s$ \citep{Poulin:2018cxd,Knox:2019rjx,kee20,Vagnozzi:2021gjh}. These two quantities are degenerate and an early phase of dark energy  can help relax the $H_0$ tension, even in the presence of the baryonic acoustic oscillations, or BAO data (see, however, the discussions in, e.g., \cite{Clark:2021hlo,Allali:2021azp} on swelling the $S_8$ tension).
 The $H_0$ tension can also be reduced by going beyond the slow roll inflation \citep{kee20}.  However, these models often fail to explain the various datasets simultaneously \citep{Jedamzik:2020zmd}.
 In an attempt for simultaneously addressing various tensions and explaining the different datasets in one model, the authors studied a phenomenological model of modified gravity  with some physical roots in the physics of critical phenomena \citep{Banihashemi:2018oxo,Banihashemi:2018has,Banihashemi:2020wtb}. 
In these works we modelled a gravitational phase transition (GPT)  at both background and linear perturbation levels by a smooth step function in redshift \cite[][referred to as the GPTs throughout this work]{Farhang:2020sij,Khosravi:2021csn}.  The degrees of freedom in the GPT scenario could simultaneously explain the Hubble and $\sigma_8$ tension, as well as the internal CMB inconsistencies of low-high $\ell$ and lensing amplitudes. 
However,  the BAO data were found to heavily push the parameters toward  their $\Lambda$CDM values and thus be in tension with direct Hubble measurements. 

In this paper, we extend the GPT model by letting the transition  be \textit{scale--dependent}. More specifically, 
the modifications to the gravity sector in perturbation equations are allowed to happen with different amplitudes at different scales. This scale-dependent gravity can be motivated by, e.g., 
physics of critical phenomena where there is a possibility to have scale--dependent forces due to nontrivial physics of microstructures.
The scale-dependency is included in the analysis in a data--driven fashion which allows for the construction of gravity eigenmodes, or GEMs. 
GEMs describe patterns of deviations from general relativity, or GR, that data are most sensitive to, and in general depend on the datasets used. 
We then investigate how the newly introduced degrees of freedom impact the tension and whether they point to deviations from GR 
through reconstructing the scale-dependent perturbation parameters.

The paper is organized as follows. We first introduce the model for gravity eigenmodes in Section~\ref{sec:model} and the data used to constrain it in Section~\ref{sec:data}. The results of the confrontation of the model with data is presented in Section~\ref{sec:results}.
We conclude the paper with discussions of the results and remarks on future work in  Section~\ref{sec:discussion}.

\section{Gravity Eigenmodes}\label{sec:model}
In the gravitational eignemode scenario, we assume that gravity undergoes a phase transition, modelled phenomenologically by a \texttt{"tanh"}, as introduced and expanded in  in the GPTs. The transition  affects both background expansion and perturbation equations, and being only a function of redshift, it is assumed to be scale--independent in the GPTs. In this paper we extend the model further to allow for scale-dependent modifications to perturbation equations. 

We assume a flat, homogenous and isotropic universe whose background evolution is described by 
\begin{eqnarray*}
	H^2(z)=H_0^2 \bigg[\Omega_r (1+z)^4+\Omega_m (1+z)^3+\Omega_\Lambda(z)\bigg].
\end{eqnarray*}
Scalar perturbations in the linear regime are introduced to the background metric through the functions $\Psi$ and $\Phi$ in the Newtonian gauge, 
\begin{eqnarray*}
        ds^2=a^2(\tau)[&-&(1+2\Psi(\tau,\vec{x}))d\tau^2 
	+(1-2\Phi(\tau,\vec{x}))d\vec{x}^2 ].
\end{eqnarray*}
The perturbations $\Psi$ and $\Phi$ must satisfy the linearly--perturbed Einstein equations. 
These equations in Fourier space can be modified phenomenologically as 
\begin{eqnarray*}\label{eq:mu}
	&&k^2 \Psi = -\mu(z,k)\,4\pi G a^2\,\left[\rho\Delta+3(\rho+P)\sigma\right],\\
	&&k^2\left[\Phi-\gamma(z,k)\Psi\right]=\mu(z,k)\,12\pi G a^2\,(\rho+P)\sigma. \label{eq:gamma}
\end{eqnarray*}
Deviations in $\mu(z,k)$ and $\gamma(z,k)$ from unity would then embrace any general redshift-- and scale--dependent  modifications to GR. 

Following the GPTs, we model the gravitational transition in the background by
\begin{eqnarray}\label{eq:delv}
	\Omega_\Lambda(z)=\Omega_{\Lambda \rm{CDM}}+\Delta_\Lambda\,\frac{1+\tanh\big[\alpha\,(z_{\rm t}-z)\big]}{2},
\end{eqnarray}
where $z_{\rm t}$ and $\alpha$ represent the redshift and inverse of the width of the transition.  
$\Omega_{\Lambda \rm{CDM}}$ corresponds to the cosmological constant in $\Lambda$CDM, so $\Delta_\Lambda=0$ would be equivalent to $\Lambda$CDM.  
  For the flat universe assumption we impose $\Omega_{\rm r}+\Omega_{\rm m}+\Omega_\Lambda(z=0)=1$, where 
$\Omega_\Lambda(z=0)=\Omega_{\Lambda \rm{CDM}}+\Delta_\Lambda[1+\tanh(\alpha z_{\rm t})]/2$. 
Similarly, for the modifications to perturbation equations, we assume a transition in $\mu$ and $\gamma$,
\begin{eqnarray}\label{eq:pert}
	\mu(z,k)&=&\mu_{\rm{GR}}+\Delta_{\mu}(k) \frac{1+\tanh\big[\alpha\,(z_{\rm t}-z)\big]}{2}, \nonumber \\
	\gamma(z,k)&=&\gamma_{\rm{GR}} + \Delta_\gamma(k) \frac{1+\tanh\big[\alpha\,(z_{\rm t}-z)\big]}{2}, 
\end{eqnarray}
where $\mu_{\rm{GR}}=\gamma_{\rm{GR}}=1$ and $\Delta_{\mu}(k)= \Delta_\gamma(k) =0$ correspond to GR.
It should be noted that in this model the redshift dependences are assumed to be the same for modifications in the background (Eq.~\ref{eq:delv}), and perturbation equations (Eq.~\ref{eq:pert}).
Scale-dependence of $\mu(z,k)$ and $\gamma(z,k)$ happens through $\Delta_{\mu}$ and $ \Delta_\gamma$ in a model-independent way.
We divide the $k$-range of interest $[k_{\rm{min}},k_{\rm{max}}]$ into $n_k$ bins.
Deviations from GR are parameterized as amplitudes of top--hats in the $k$-bins of $\mu$ and $\gamma$, i.e., through non-zero 
$\Delta_{\mu_i}$ and $\Delta_{\gamma_i}$,  
$i=1,.., n_k$. 
Therefore, in addition to the standard $\Lambda$CDM parameters, the parameter space of this phenomenological model of cosmology would consist of $\{z_{\rm t}, \alpha, \Delta_\Lambda, \Delta_{\mu_1}, ..., \Delta_{\mu_{n_k}}, \Delta_{\gamma_1}, ..., \Delta_{\gamma_{n_k}}\}$ (see Table~\ref{tab:params}).
We modify the code MGCAMB\footnote{https://github.com/sfu-cosmo/MGCAMB. \\
MGCAMB is an implementation of  CAMB, https://camb.info/,  for modified gravity models.}\citep{Hojjati:2011ix}, to take into account our desired changes.
The new parameters are then fed into the Cosmological MonteCarlo code, CosmoMC\footnote{https://cosmologist.info/cosmomc/}, to be measured with various dataset combinations. 

In general, the correlation between these perturbation parameters could be large, in particular for adjacent bins, due
to similar traces they leave on data. On the other hand, adding  $\sim 2n_k$ new parameters is a quite expensive approach to describe possible deviations around the astonishingly simple $\Lambda$CDM.
One could therefore consider constructing $2 n_k$ uncorrelated linear combinations of  $\Delta_{\mu_i}$'s and $\Delta_{\gamma_i}$'s 
and keeping only the most tightly constrained ones. The new parameters are generated through the eigen-decomposition of the perturbation blocks of the covariance matrix, $\bf{C}_\mu$ and ${\bf C}_\gamma$. They are often referred to as the principal components or eigenmodes of the covariance matrix, hence the name gravity eigenmodes.  Estimates of their errors are given by the square roots of the eigenvalues of $\bf{C}_\mu$ and ${\bf C}_\gamma$.  
The eigenmodes and their measured means and estimated uncertainties can then be used to reconstruct the overall patterns of observed deviation from unity for $\mu$ and $\gamma$, at scales of interest.

\begin{table}\centering
    \begin{tabular}{cc}
    \hline  \hline 
    parameter & description \\  \hline
    $\Delta_\Lambda$ & transition amplitude for $\Omega_\Lambda$ \\
    $z_{\rm t}$ & transition redshift\\
    $\alpha$ & inverse of the width of transition\\
    $\Delta_{\mu_i}$  & transition amplitude for $\mu$,  $i \in \{1,...,n_k\}$  \\
    $\Delta_{\gamma_i}$ &  transition amplitude for $\gamma$,  $i \in \{1,...,n_k\}$\\
  \hline
\end{tabular}
\caption{Model parameters.}
\label{tab:params}
\end{table}
\section{datasets}\label{sec:data}
In this work we use {\it Planck} CMB measurements of  temperature and polarization anisotropies and its lensing, referred to as P18 \citep{pl18}, the local measurements of the Hubble constant, labeled  as R21 \citep{Riess:2021jrx}, BAO measurements  consisting of 6DFGS \citep{Beutler:2011hx}, SDSS MGS \citep{Ross:2014qpa} and BOSS DR12 \citep{Alam:2016hwk},  and the measurements of cosmic shear and galaxy clustering by Dark Energy Survey, referred to as DES \citep{Kazin:2014qga}. 
We also consider a mock futuristic and extremely high--precision measurement of the Hubble constant with $H_0=73.30 \pm 0.01$, to investigate the implications of such a measurement for the construction of $\Delta_{\mu}(k)$ and $\Delta_{\gamma}(k)$. We refer to this Hubble measurement as $H_073$ for short.  

\section{results}\label{sec:results}
\begin{table*}\centering
    \begin{tabular}{ccccccccc}
    \hline  \hline 
    & $\Omega_{\rm b}h^2$ &  $\Omega_{\rm c}h^2$ & $H_0$ & $\tau$ & $\ln (10^{10}A_{\rm s})$ & $n_{\rm s}$ \\\hline 
    P18+& & GEM&& &   \\ \hline 
   BAO+R21&$0.0224\pm 0.0003 $ & $0.119\pm  0.002$ & $70.3\pm  6.8   $ & $0.053  \pm 0.007$ &  $3.040\pm  0.016 $& $0.967\pm  0.007$  \\ 
   BAO&$0.0225\pm  0.0002 $ & $0.119\pm  0.002$ & $71.9\pm  9.2  $ & $0.053  \pm 0.008$ &  $3.037\pm 0.017$& $0.969\pm  0.006$   \\ 
   R21&$0.0225\pm  0.0002 $ & $0.119\pm  0.002$ & $72.3 \pm 2.6 $ & $0.054\pm 0.009$ &  $3.039\pm  0.017$& $0.967\pm  0.005$   \\   \hline
 & &$\Lambda$CDM & \\ \hline 
 BAO & $ 0.0224 \pm 0.0001$ &   $ 0.119 \pm 0.001$& $67.7 \pm 0.4$  & $0.0561 \pm 0.0071$ & $3.047 \pm 0.014 $ & $0.967 \pm 0.004$\\
  \hline
\end{tabular}
\caption{Standard cosmological parameters in GEM and $\Lambda$CDM measured by various datasets.}
\label{tab:std}
\end{table*}
The goal is to find  how  small deviations from GR, parametrized in a semi-blind way, would alleviate some of the cosmological tensions. 
We take $n_k=20$  logarithmically--spaced bins in the range $[k_{\rm{min}},k_{\rm{max}}]=[10^{-5},1]$. 
For the the redshift of transition $z_{\rm t}$ we first search in $ [10^{-4}, 10^4]$ to cover a large redshift range of cosmological interest. We explore this range logarithmically for a relatively fair coverage of different redshift scales.
 The transition is also allowed to occur both ways, from $\Lambda$CDM  at late times to the perturbed model at early times and vice versa by allowing $\alpha$ to be positive as well as negative.
However, we find no support from data for such a transition at high $z$. 
We therefore limit the search to a linear search in the range $z\in[10^{-4},20]$ for higher resolution.
We are  interested to see whether the Hubble tension between R21 and P18+BAO datasets and  the $\sigma_8$ tension between P18 and DES are reduced in this scenario.
\begin{figure}
 \begin{center}
 \includegraphics[scale=0.5]{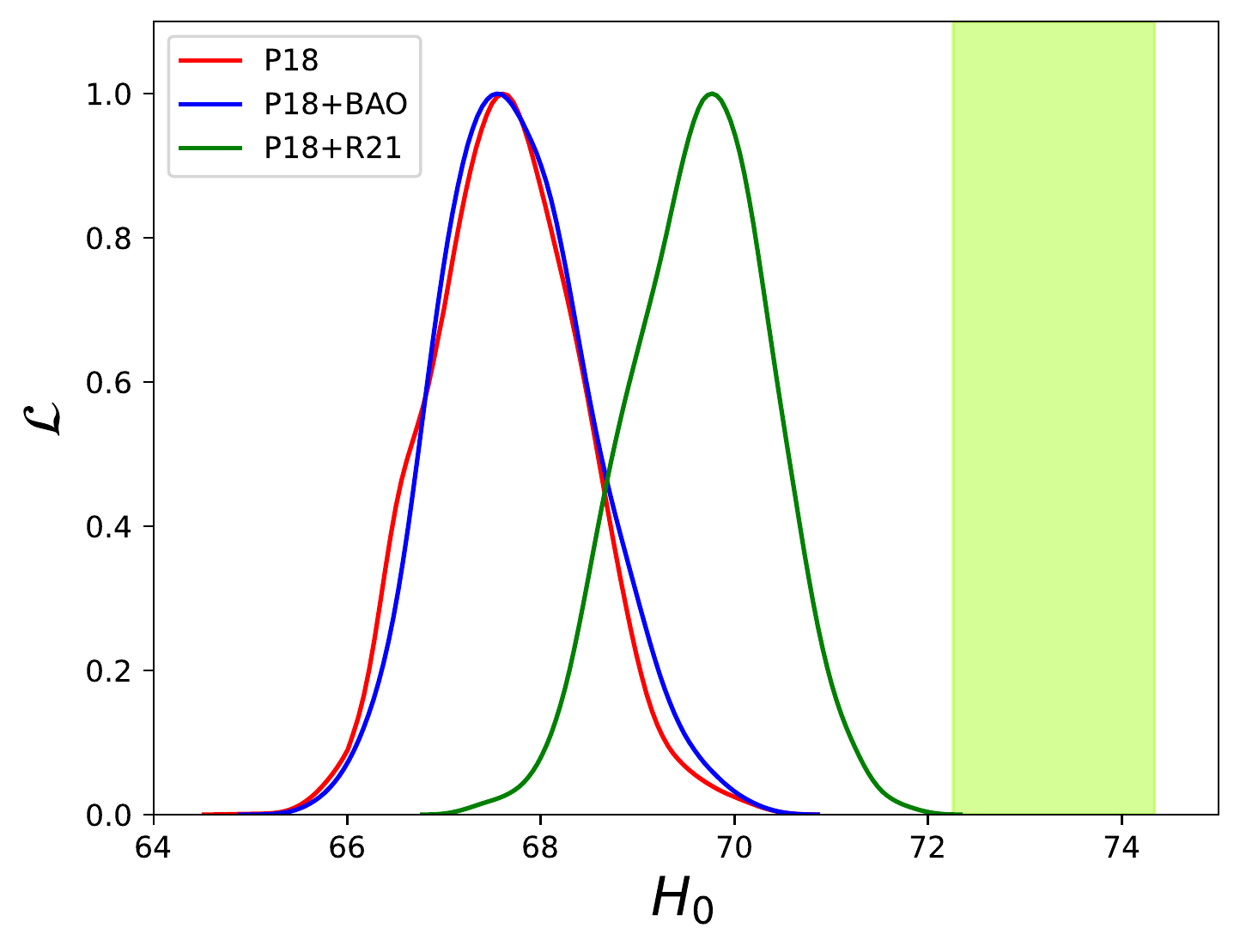}    
 \includegraphics[scale=0.5]{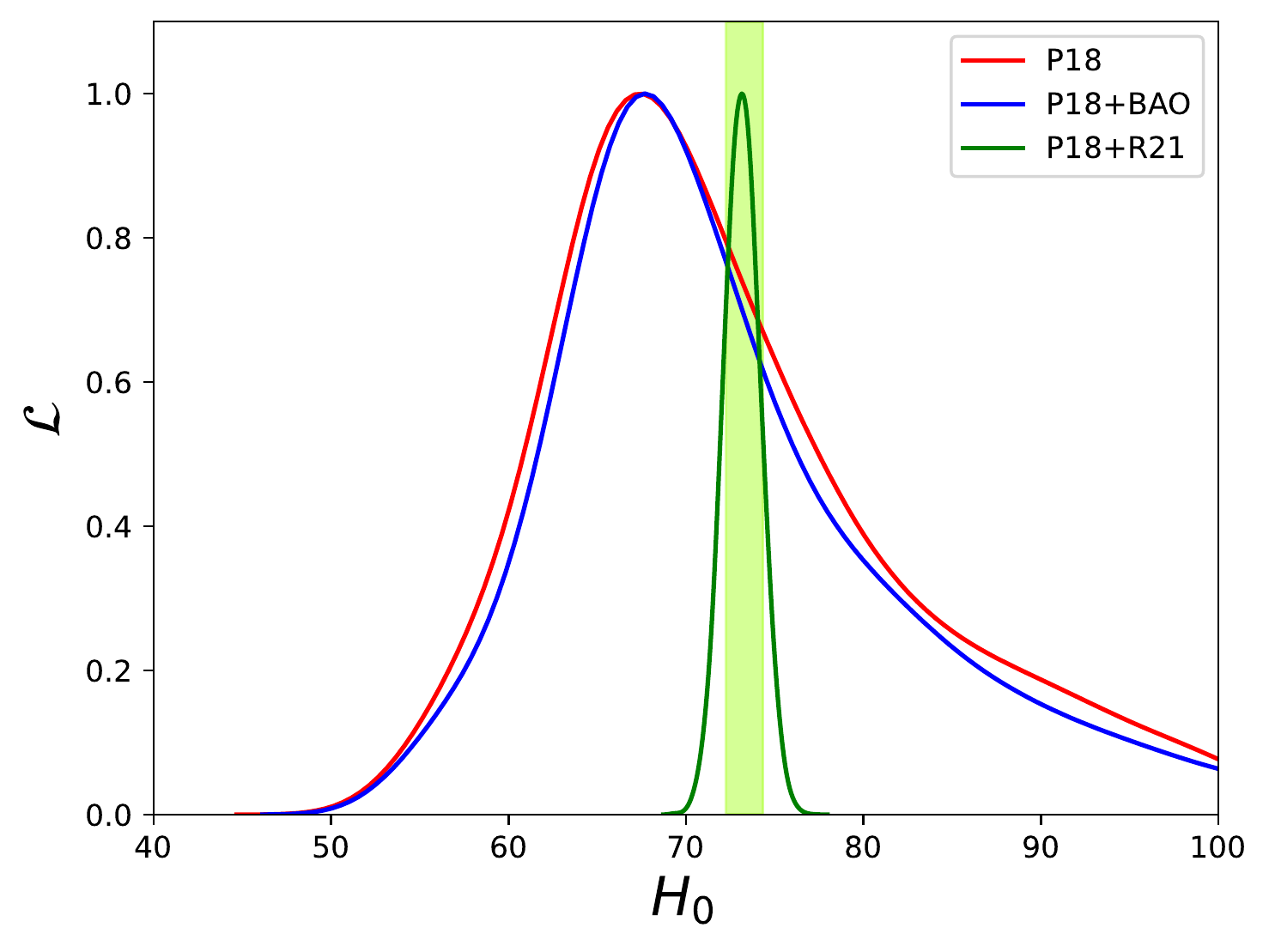}    
   \caption{The posterior probability distribution of $H_0$ for three data combinations. 
   The shaded green regions illustrate the R21 measurement of $H_0$.
   Top: The background evolution is assumed to be the same as in  $\Lambda$CDM. Bottom: The background evolves according to Eq.~\ref{eq:delv}. }
 \label{fig:H0}
 \end{center}
 \end{figure}

\subsection{Hubble tension}
We first investigate how various combinations of P18, BAO and R21  constrain standard and GEM parameters. 
We compare the results for two cases, with and without modifications in the background evolution. This would help  better 
distinguish the impact of $\mu$ and $\gamma$ perturbations  from background modifications on observables.
%
\begin{figure*}
 \begin{center}
 \includegraphics[scale=0.37]{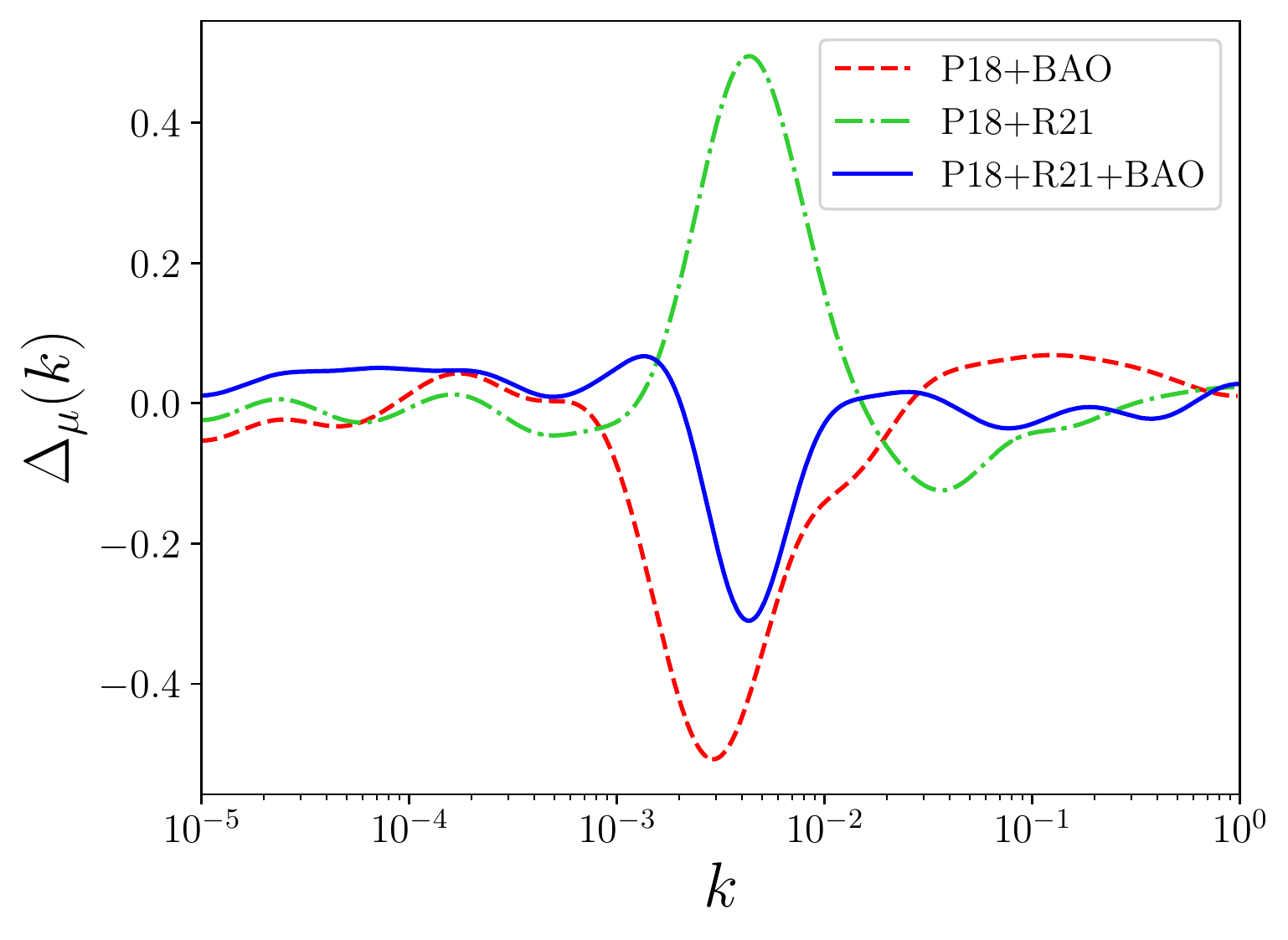}    
 \includegraphics[scale=0.37]{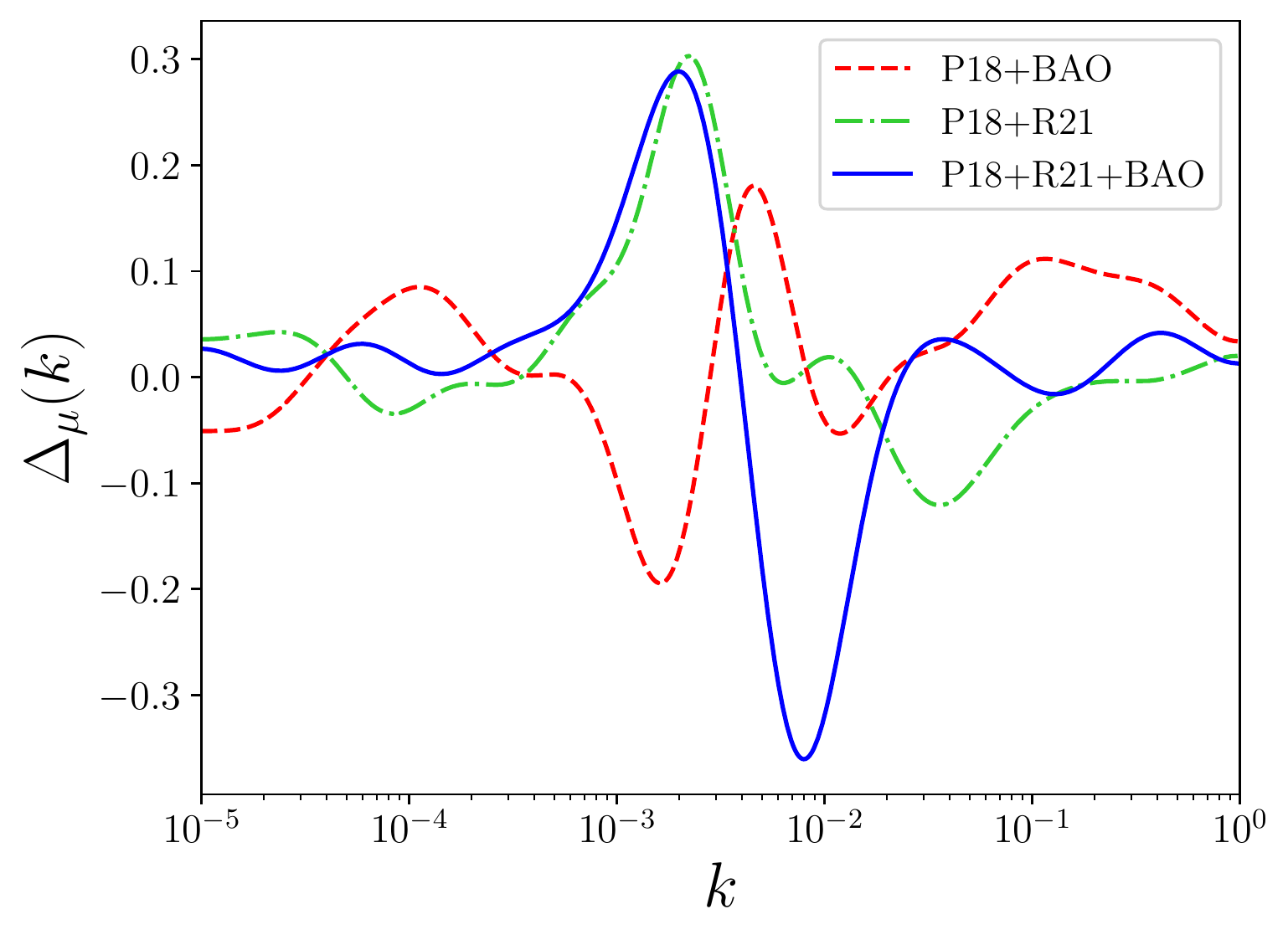}    
 \includegraphics[scale=0.37]{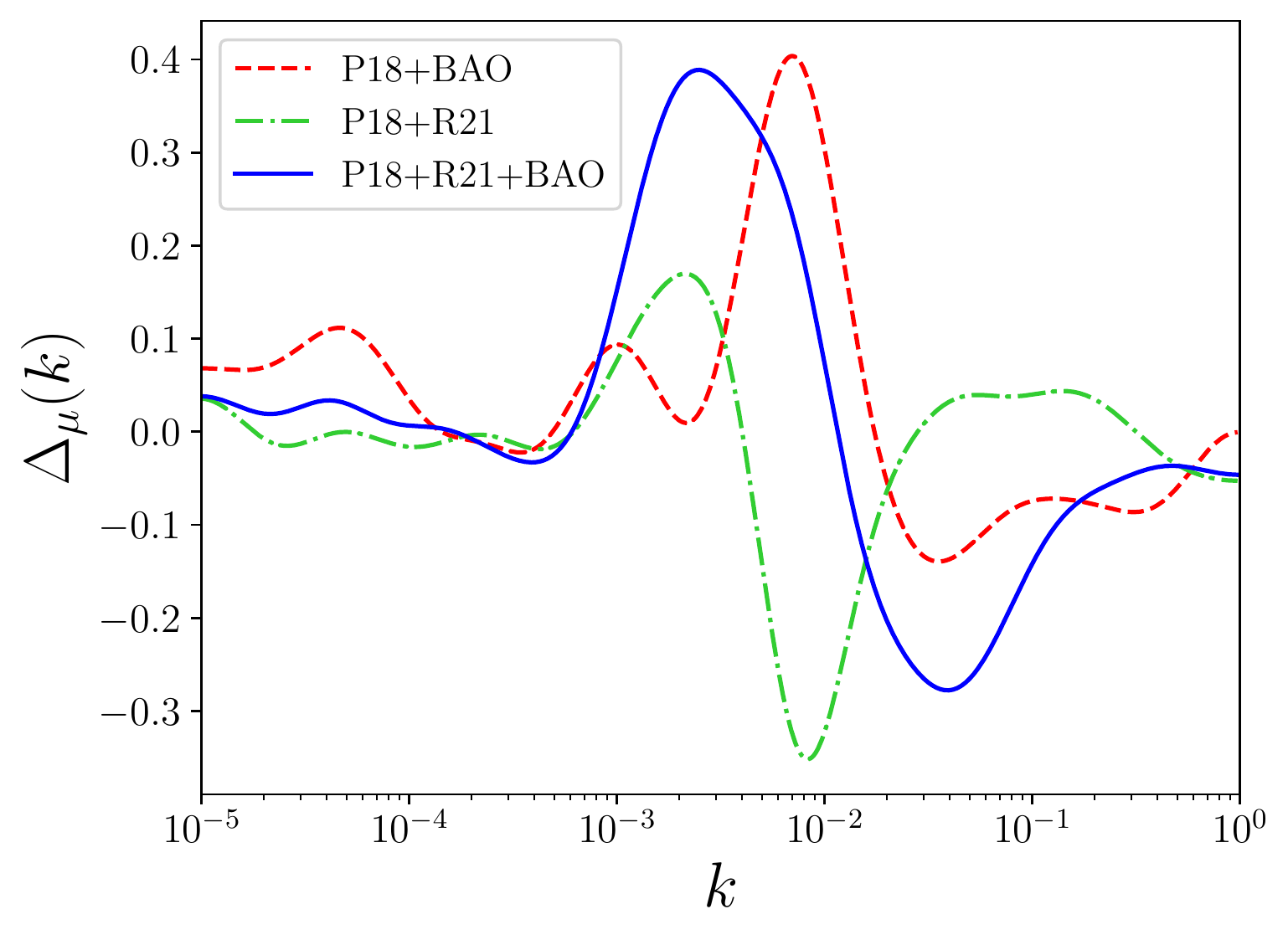}  
 \includegraphics[scale=0.37]{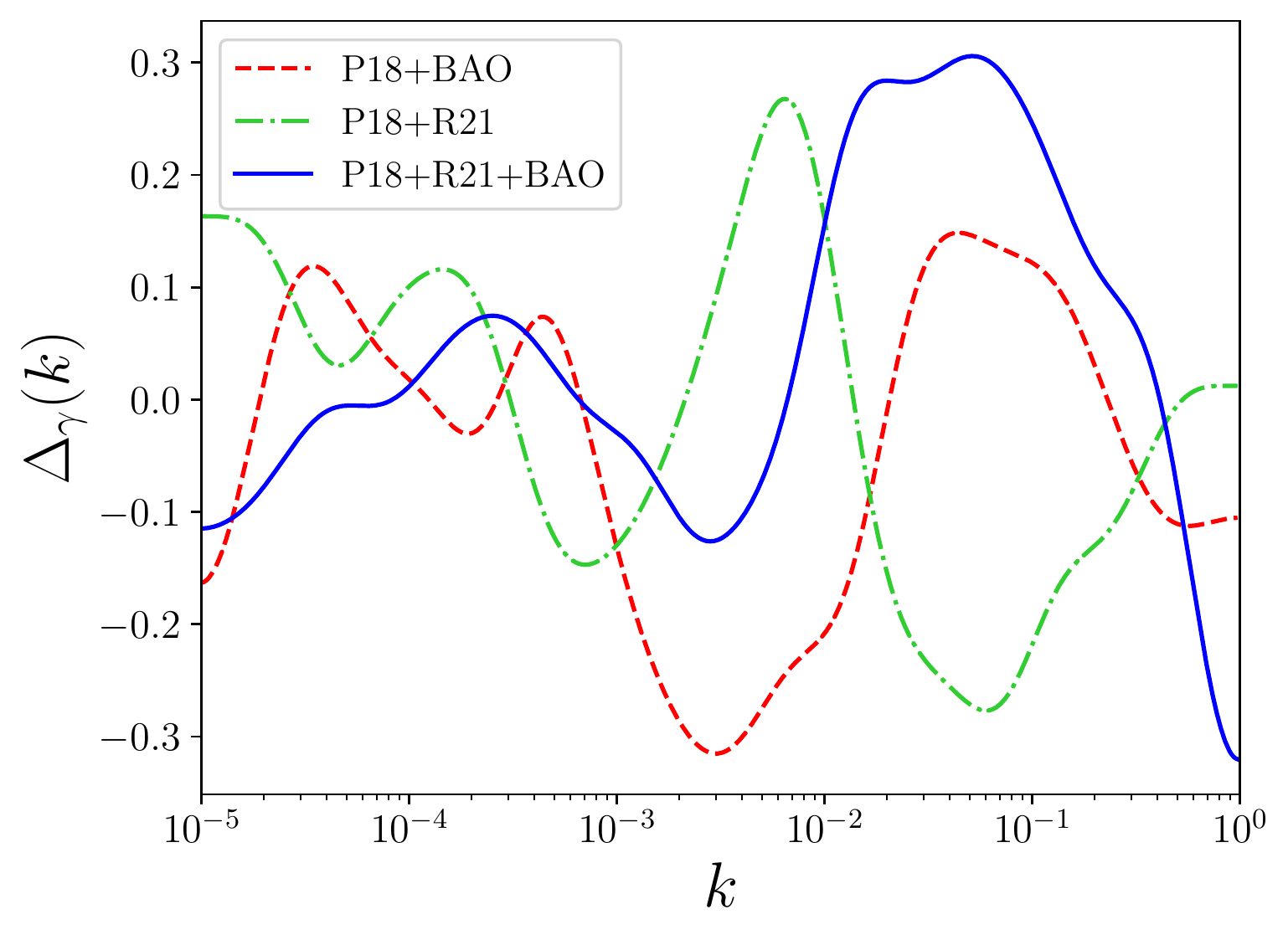}    
 \includegraphics[scale=0.37]{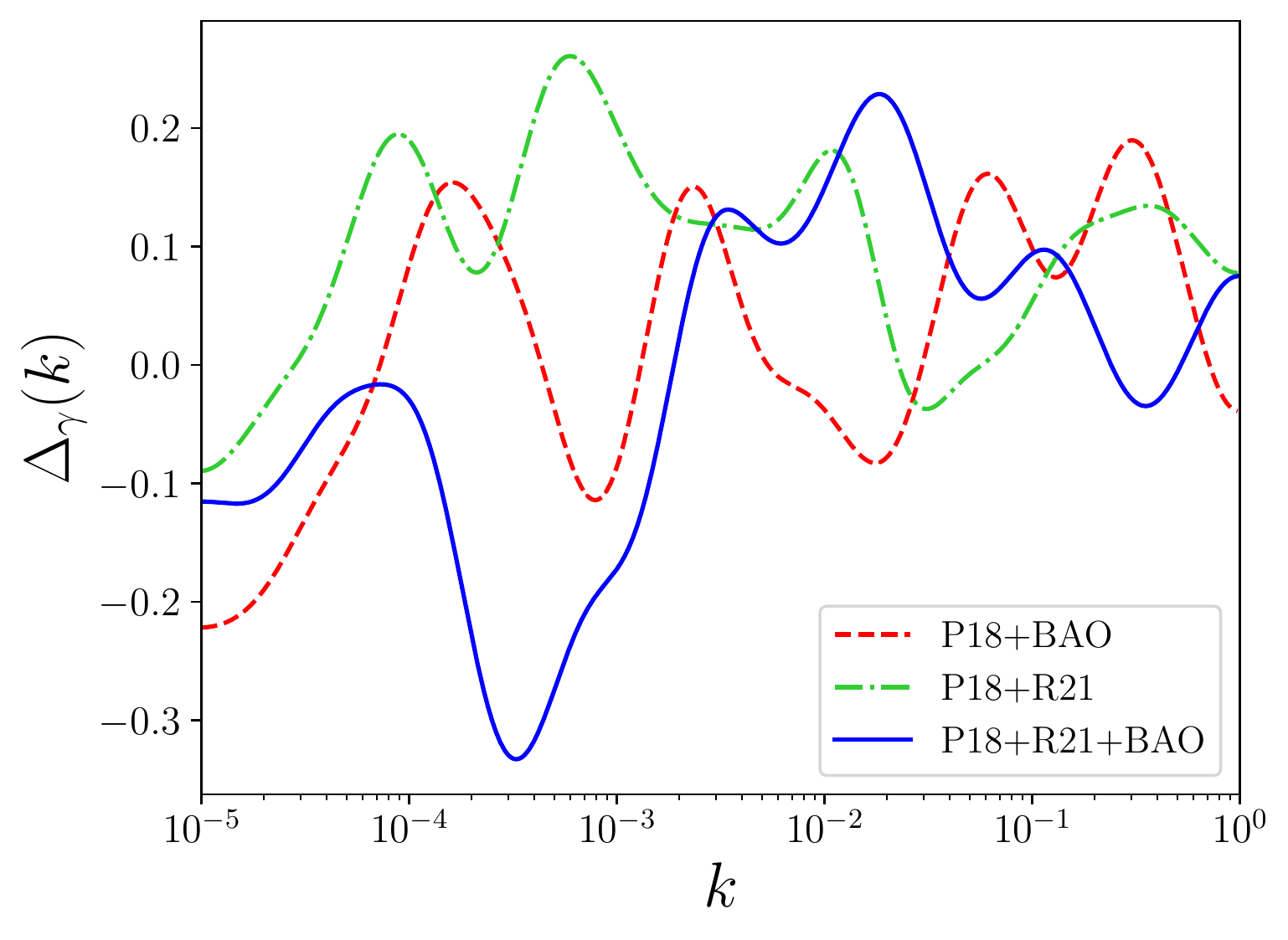}    
 \includegraphics[scale=0.37]{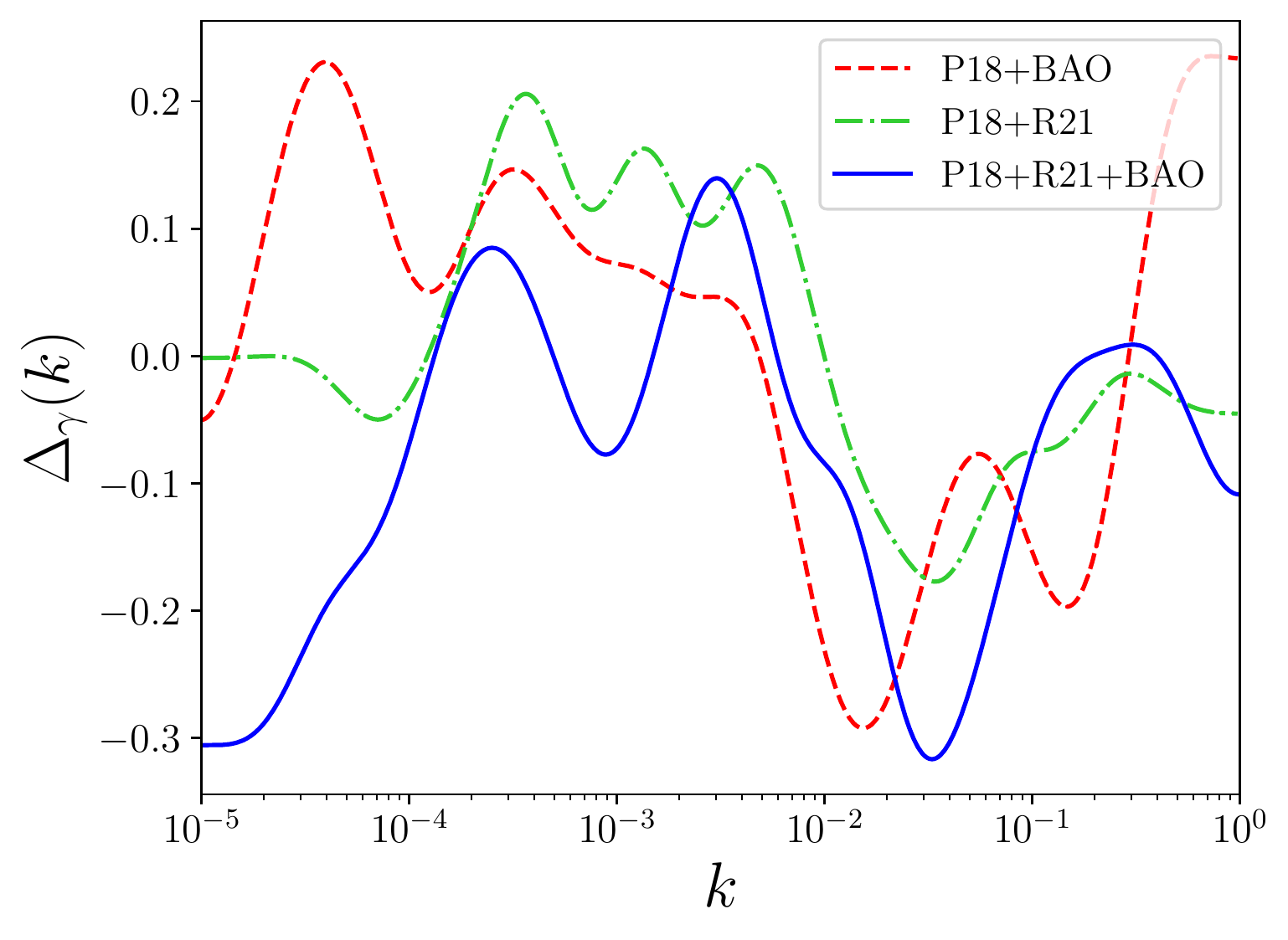}    
\caption{The first three GEMs, or $k$-dependent gravity eigenmodes, for $\mu$ and $\gamma$, with various datasets. The background is allowed to experience a phase transition as in Eq.~\ref{eq:delv}.}
 \label{fig:gem:p}
 \end{center}
 \end{figure*}

Figure~\ref{fig:H0} illustrates the posterior probability of $H_0$ for these two cases.
In both plots the likelihoods for P18 and P18+BAO agree, confirming the consistency of the two datasets as far as $H_0$ is concerned.
 It is, however, evident from the top panel that modifications in $\mu$ and $\gamma$ alone, i.e., with the background assumed to be the same as $\Lambda$CDM,  cannot resolve the Hubble tension between P18 (or P18+BAO) and R21, and there still exists a $\sim 2-3 \sigma$ tension between the two measurements. 
The joint P18+R21 likelihood, despite the inconsistency of P18 and R21 in this scenario, is plotted to see how far P18 would push the bestfit $H_0$ of R21 toward its preferred lower values.  Allowing the background to also deviate from $\Lambda$CDM would largely increase the uncertainty in the $H_0$ measurement,
and let $H_0$ adopt higher values even with P18,
 as shown in the bottom panel.
 With this enhanced uncertainty of  $H_0$ measurement,  the Hubble tension  is fully removed. 
It is interesting to note that the joint P18+R21 measurement of $H_0$, which are now justifiably combined, yields a quite large bestfit for $H_0$, unlike the P18+R21 measurement in the fixed-background scenario. 
Also note that background--only modifications, and with no new degrees of freedom in perturbation equations, would not relax the Hubble tension, as was explored and fully discussed in \cite{Khosravi:2021csn}.  

 Table~\ref{tab:std} compares the mean values and errors for the standard cosmological parameters in GEM and  $\Lambda$CDM scenarios, for various dataset combinations. Note that the datasets R21 and P18+BAO are incompatible in $\Lambda$CDM  and cannot be combined.  
Moreover, we find that $\Delta_\Lambda$ is basically unconstrained and  hits the prior boundaries $[-1,1]$ for all data combinations used here.
The transition is  preferred to occur at $z_t < 1$ and in both directions (determined by the sign of $\alpha$).
 However, a transition from $\Lambda$CDM at high redshifts to GEM at lower redshifts is slightly  favoured by data. 
We note that although the degrees of freedom in $\mu$ and $\gamma$ are  important to relax the tension, these parameters are found to be consistent with GR. 

The obscuration of possible deviation in the new parameters from their GR  and $\Lambda$CDM values could be explained by the large number of perturbation parameters, as
discussed in Section~\ref{sec:model}.  
We therefore use the $n_k \times n_k$ correlation matrices of perturbation amplitudes $\bf{C}_\mu$ and  $\bf{C}_\gamma$ to generate the gravity eigenmodes. The amplitudes of the modes and their errors can be calculated respectively from the measurements of the top--hat amplitudes and their covariance matrices.
 This method of eigenmode construction is data--driven in the sense that the modes can be rank--ordered so that the first few ones represent perturbation patterns with most detectable imprints on data. 
 Therefore one expects smaller errors on the amplitudes of these modes, compared to the top--hats we started with, and a higher chance of detecting semi--blind, scale--dependent deviations from GR.

The first three perturbation eigenmodes of $\Delta_\mu(k)$ and $\Delta_\gamma(k)$ are presented in Figure~\ref{fig:gem:p}. The $\Delta_\mu$  eigenmodes have their main features in $k \in \{0.001, 0.1\}$, extending to higher $k$'s as the mode number increases. The modes for the full data combination P18+BAO+R21 are the smoothest with fewer wiggles, and constructed with fewer, more dominant bumps.  
For $\gamma$ the case is different and the modes seem to be extended in a large $k$-range.
It should be stated that for some $\Delta_{\gamma_i}$'s, unlike $\Delta_{\mu_i}$'s, the errors are dominated by the prior ranges. On the other hand, the priors cannot be increased farther due to numerical limitations. This would make the patterns of  $\gamma$ eigenmodes  affected by noise and numerical insufficiencies.

\begin{figure}
 \begin{center}
 \includegraphics[scale=0.5]{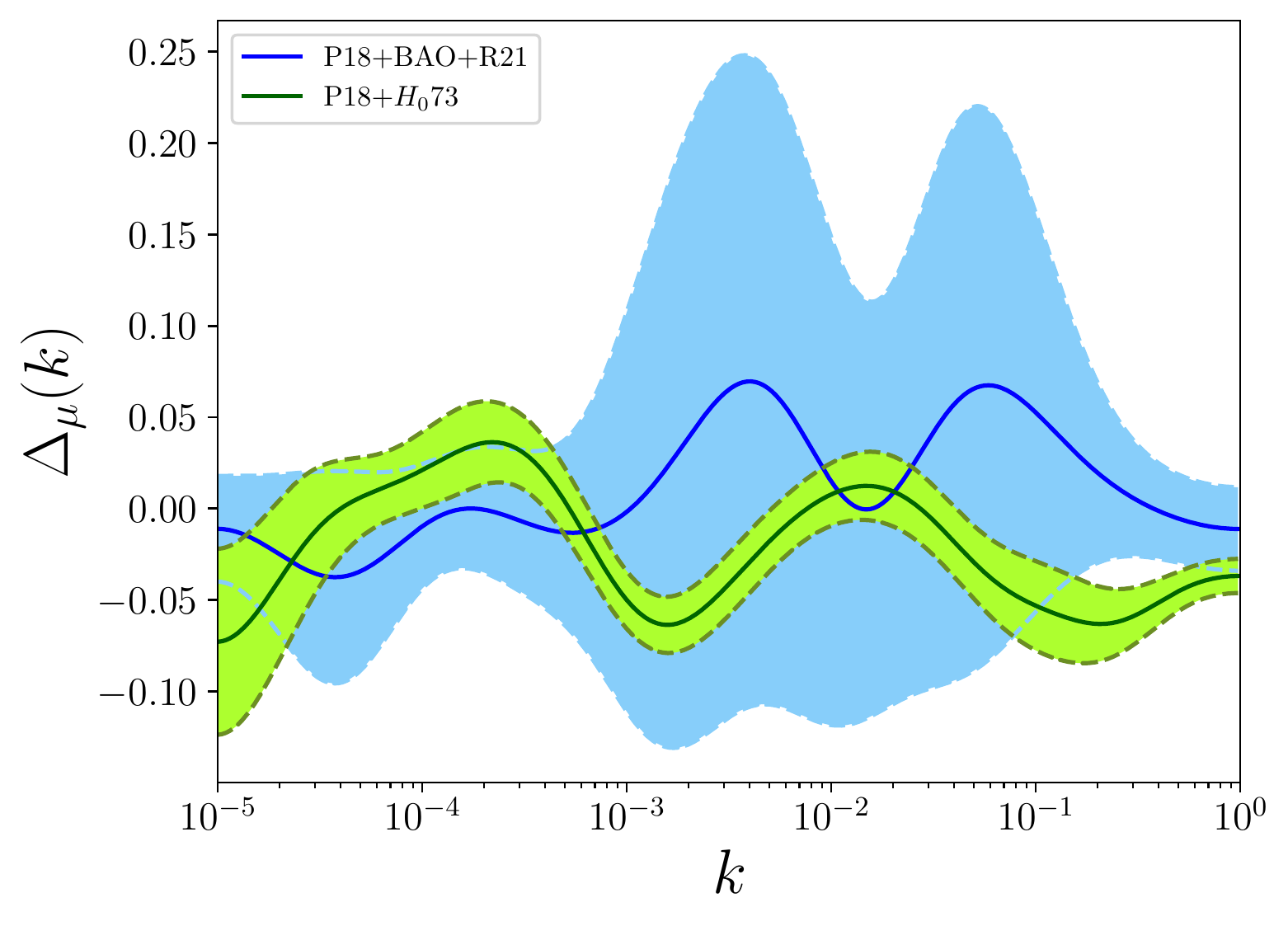}    
 \includegraphics[scale=0.5]{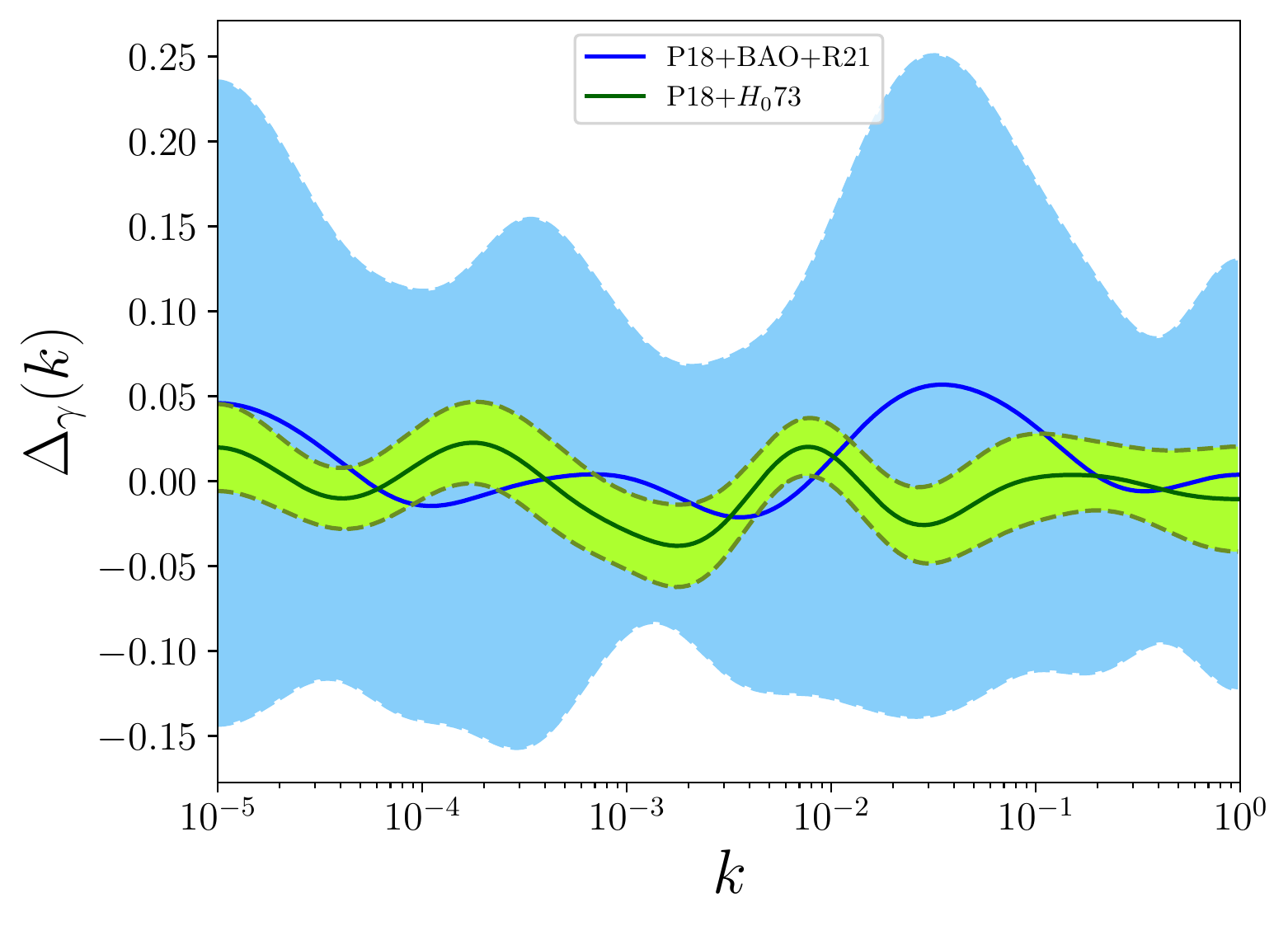}    
   \caption{The reconstructed perturbations in $\mu(k)$  and $\gamma(k)$  in the GEM scenario, with the first five eigenmodes, for the two data combinations P18+BAO+R21 and P18+$H_073$.}
 \label{fig:recons}
 \end{center}
 \end{figure}

The estimated errors for these modes are listed in Table~\ref{tab:p:sigma}. All the mode amplitudes are found to be consistent with zero, and therefore no hint $\ge 1\sigma$ is found for deviation from GR.   
However, it is important to stress that the Hubble tension is fully resolved in our  phenomenological extension of GR.  %
It is interesting to see the $\Delta_\mu(k)$ and $\Delta_\gamma(k)$ patterns leading to this resolution. We reconstruct these perturbation patterns using the first five eigenmodes. We use the linear dependance of the modes on the $\mu$ and $\gamma$ top-hats, and their measured amplitudes to infer the means and errors of the GEMs. 
The resulting patterns of deviations are shown in Figure~\ref{fig:recons} as solid blue curves. 
The shaded blue areas around the mean patterns illustrate the region of $1\sigma$ uncertainty and are found based on the estimated errors of the mode amplitudes and their pairwise independence. 
Therefore the price to resolve the Hubble tension in this model has been the (insignificant) deviations from GR in $\mu$ and $\gamma$, mainly for $k \in \{0.001, 0.1\}$.
The fact that GEM can resolve the Hubble tension, yet  no deviations from $\Lambda$CDM are detected, persuades one to explore how a very tight $H_0$ measurement, in substantial tension with early $\Lambda$CDM-inferred measurements of the Hubble constant, would derive GEM parameters away from their GR values.  
The green solid lines and shaded areas in Figure~\ref{fig:recons} illustrate the means and $1\sigma$ regions of uncertainty for the  reconstructed $\Delta_\mu(k)$ and $\Delta_\gamma(k)$  patterns with this $H_073$ mock dataset (see Section~\ref{sec:data}). 
To compensate for the tightly measured high value of $H_0$, $\mu$ and $\gamma$ need to differ from $1$ at more than  $1\sigma$ at various scales.  With this mock dataset we also find a higher than $5\sigma$ detection of $\Delta_\Lambda$ with $\Delta_\Lambda=0.44 \pm 0.08$, and with a transition at $z_{\rm t}=0.14 \pm 0.11$ with $\alpha=2.2\pm 0.8$.

\subsection{ Tension in $\sigma_8$}
We use the data from DES to constrain parameters in the GEM scenario and investigate how they compare to P18 predictions.  
The measurements of $\sigma_8$ are of prime interest in this case as they happen to show a mild, $\sim 2\sigma$ disagreement  in the $\Lambda$CDM framework for P18 and DES. 
For  the DES case we have  imposed a relatively large prior on $H_0$, i.e., $H_0 \in [55,95]$.
In the GEM scenario we find 
\begin{eqnarray}
\sigma_8=0.78\pm  0.11  ~~~ {\rm DES}\\
\sigma_8=0.85\pm 0.07  ~~~~ {\rm P18}
\end{eqnarray}
showing that the $\sigma_8$ tension is reduced to less than one sigma.

\begin{table}\centering
    \begin{tabular}{ccccccc}
    \hline  \hline 
    & $\sigma_{\mu_1}$ &  $\sigma_{\mu_2}$ & $\sigma_{\mu_3}$ & $\sigma_{\gamma_1}$ & $\sigma_{\gamma_2}$ & $\sigma_{\gamma_3}$ \\ \hline 
  P18+BAO&$ 0.38  $&$  0.40 $&$  0.43 $& $  0.45$&$ 0.47 $&$ 0.49  $\\
  P18+R21& $ 0.35$&$  0.37 $&$  0.38  $&$  0.46 $&$ 0.46$&$0.47$ \\
  P18+R21+BAO &  $ 0.34 $&$  0.35 $&$  0.37 $&$  0.44$&$ 0.45  $&$   0.46 $\\
  P18+$H_0$73 & 0.04 & 0.04 & 0.07 &  0.04 & 0.06 & 0.07 \\
  \hline
\end{tabular}
\caption{The estimated uncertainty of the eigenmodes plotted in Figure~\ref{fig:gem:p}. }
\label{tab:p:sigma}
\end{table}
\section{Discussion}\label{sec:discussion}
In this work we explored a phenomenological extension to general relativity, with perturbation equations allowed to deviate from  GR  in a scale--dependent way. 
Based on the two earlier GPT works,  this deviation was assumed to accompany a phase transition in redshift, both at the background and perturbation levels. 
The scale-dependency was investigated in a semi--blind way by generating gravity eigen-modes, as patterns of modifications to GR with the highest impact on data. 
 Our results showed that the enhanced parameter space could consistently accommodate  CMB, BAO,  local Hubble measurements and data from large scale structure. However, it should be noted that no significant departure from $\Lambda$CDM and GR was observed.

 To test the model against a more restrictive scenario, we assumed a mock measurement of the Hubble constant $H_0=73.30\pm 0.01$.
 This led to tighter constraints on our free parameters. 
 In particular, our forecasts showed the inconsistency of $\Delta_\mu(k)$, which represents modification to Poisson equation, from GR predictions if the Hubble constant insisted on the large value $H_0=73.30$ (see Figure \ref{fig:recons}).
 On the other hand, $\Delta_\gamma(k)$ which modifies the lensing potential, did not show any significant departure from GR. 
 We therefore expect future data with their higher resolution to be able to distinguish the GEM model from GR, with tighter constraints on GEM parameters. 
 The $k$--dependent freedom in $\mu$ and $\gamma$ amplitudes may also have the potential to relax the low-high $\ell$ and lensing tensions in CMB. We leave this to future work.  

The $k$-reconstruction is a plausible approach to look for modified gravity signatures. 
However, introducing many degrees of freedom to the extremely simple and neat $\Lambda$CDM  is a quite high price to relax the tensions. Yet, it should be noted that this could be an issue for a theoretical model, while the GEM scenario is 
phenomenological in redshift space and data--driven in $k$-space.
For such a theoretically--blind approach, the ultimate goal could be adding as many degrees of freedom as numerically possible to allow the data to freely choose the ones best describing it. 
This is simply not feasible due to limited available information in the data and computational power. 
Hence one can cut the mode--hierarchy by keeping the few modes with the tightest constraints and leave
the rest out of the analysis.

It should also be noted that this search for non-trivial $k$-dependency could be motivated by the physics of critical phenomena where it is possible for the system to behave differently in different scales, associated with a transition in time. That is why we restricted the gravitational modification in this work to be accompanied with a phase transition. 
This assumption can however be relaxed to explore a more general reconstruction scenario of gravity in  redshift and scale, without assuming a restrictive theoretical prior \cite[see][for reconstructed modifications of gravity in time]{Raveri:2019mxg}.
This approach would suffer from the many degrees of freedom introduced to cover possible modification in large spans of redshifts and scales. It is therefore limited by numerical insufficiencies, and calls for a clever way to look at. We leave this to future work. 

\section{Acknowledgement}
Part of the numerical computation of this work was carried out on the computing cluster of the Canadian Institute for Theoretical Astrophysics (CITA), University of Toronto.

\bibliography{gem}
\bibliographystyle{aasjournal}

\end{document}